# Anisotropic Magnetotransport and Exotic Longitudinal Linear Magnetoresistance in $WTe_2$ Crystals

Yanfei Zhao[1,2], Haiwen Liu[1,2], Jiaqiang Yan[3,4], Wei An[2,5], Jun Liu[6], Xi Zhang[1,2], Hua Jiang[7], Qing Li[8], Yong Wang[6], Xin-Zheng Li[2,5], David Mandrus[3,4], X. C. Xie[1,2], Minghu Pan[9,*] and Jian Wang[1,2]*

[1] International Center for Quantum Materials, School of Physics, Peking University, Beijing 100871, China
[2] Collaborative Innovation Center of Quantum Matter, Beijing 100871, China
[3] Department of Materials Science and Engineering, University of Tennessee, Knoxville, Tennessee 37996, USA.
[4] Materials Science and Technology Division, Oak Ridge National Laboratory, Oak Ridge, Tennessee 37831, USA.
[5] School of Physics, Peking University, Beijing 100871, China
[6] Center of Electron Microscopy, State Key Laboratory of Silicon Materials, Department of Materials Science and Engineering, Zhejiang University, Hangzhou, 310027, China
[7] College of Physics, Optoelectronics and Energy, Soochow University, Suzhou 215006, China
[8] Institute of Functional Nano and Soft Materials (FUNSOM) and Collaborative Innovation Center of Suzhou Science and Technology, Soochow University, Jiangsu 215123, China.
[9] School of Physics, Huazhong University of Science and Technology, Wuhan 430074, China.

*Correspondence and requests for materials should be addressed to M. P. (email: mhupan @gmail.com) or to J. W. (email: jianwangphysics@pku.edu.cn)

$WTe_2$ semimetal, as a typical layered transition-metal dichalcogenide, has recently attracted much attention due to the extremely large, non-saturating parabolic magnetoresistance in perpendicular field. Here, we report a systematic study of the angular dependence of the magnetoresistance in $WTe_2$ single crystal. The violation of the Kohler rule and a significant anisotropic magnetotransport behavior in different magnetic field directions are observed. Surprisingly, when the applied field is parallel to the tungsten chains of $WTe_2$, an exotic large longitudinal linear magnetoresistance as high as 1200% at 15 T and 2 K is identified. Violation of the Kohler rule in transverse magnetoresistance can be understood based on a dual effect of the excitons formation and thermal activation, while large longitudinal linear magnetoresistance reflects perfectly the scattering and nesting of quasi-1D nature of this balanced hole-electron system. Our work will stimulate studies of such double-carrier correlated material and corresponding quantum physics.

Contrast to the classical quadratic magnetoresistance (MR) in metals and semiconductors, linear magnetoresistance (LMR) is an unusual phenomenon in condensed matters. LMR has been found only in limited materials, including silver chalcogenides [1], multilayer epitaxial graphene [2], topological insulators [3-5], three dimensional Dirac semimetal such as $Cd_3As_2$ [6] and TlBiSSe [7], which has often been interpreted by either the theory proposed by Abrikosov [8] or the inhomogeneity in materials [1]. LMR behaviors are usually observed in a magnetic field applied perpendicular to the direction of current and have become one of the significant research directions in condensed matter physics and material science.

Tungsten ditelluride ($WTe_2$) semimetal, due to its extremely large, non-saturating magnetoresistance (XMR) has gained significant research interest recently [9-14]. $WTe_2$ is a typical layered transition-metal dichalcogenide (TMD) with tungsten chains sandwiched by adjacent chalcogenide layers. The extremely small overlap between the bottom of conduction band and the top of valance band in $WTe_2$ semimetal results in many interesting properties, such as the complicated band structure with multiple Fermi pockets [9, 10], the XMR effect reaching $1.3\times 10^7$ % at 60 T and 0.53 K in perpendicular field [9], as well as the pressure induced superconductivity[13, 14]. However, detailed angle-dependent MR study, representing anisotropic properties of $WTe_2$, has not been fully investigated and still remains unclear.

In this letter, we report a systematic study of the angular dependence of the magnetoresistance in $WTe_2$ single crystal, as well as a function of temperature. A large residual resistance ratio (RRR) of 741 in zero-field and a significant anisotropic magnetotransport behavior in different magnetic field directions are observed. Strikingly, when the applied magnetic field is parallel to the tungsten chains (*a* axis) of $WTe_2$, an exotic large longitudinal LMR as high as 1200% at 15 T and 2 K is identified. Our results suggest that, due to its balanced hole and electron populations, $WTe_2$ semimetal seems to be the first known material for that the longitudinal LMR appears when the magnetic field is applied parallel to the applied current.

W and Te shots in an atomic ratio of 1:49 were placed in a 5 ml $Al_2O_3$ crucible to grow the $WTe_2$ single crystals. A catch crucible containing quartz wool was mounted on top of growth crucible and both were sealed in a silica ampoule under approximately 1/3 atmosphere of high pure argon gas. The sealed ampoule was heated up initially to 1100℃ and kept for 6 hours, then cooled down to 500℃ over 96 hours. The Te flux was separated from single crystals by using a centrifuge once the temperature reaches 500℃. The $WTe_2$ single crystal grown by this method, further studied by FEI Tecnai F20 transmission electron microscope operated at 200 kV, is proven to be of high quality. Figure 1(a) displays the atomically high-resolution transmission electron microscopy (HRTEM) image, suggesting the crystal growth preferentially along the [001] direction (*c* axis). Figure 1(b) shows the selected area electron diffraction (SAED) pattern looking down the [100] zone, further indicating the single crystal structure of $WTe_2$.

An optical image of a typical $WTe_2$ single crystal for the transport measurement is shown in the inset of Fig. 1(c). The transport results reported here are primarily from the sample with the size of ∼ 1.5 mm × 0.2 mm × 15 μm (length × width × thickness). The observations are further confirmed by measurements on other $WTe_2$ samples from the same batch. Standard four-electrode method was used for the measurement with the current along the tungsten chains (*a* axis). Figure 1(c) shows the resistivity as a function of temperature at different magnetic fields, which are perpendicular to the chalcogenide layers (along the *c* axis). With decreasing temperature, the resistivity exhibits a well-metallic behavior at zero-field by showing a huge residual resistance ratio (RRR = 741), much larger than the reported values in other semimetals such as Bi [15] and $NbSb_2$ [16]. Interestingly, when applying the magnetic field, the resistivity displays a remarkable increase in low temperature regime showing a metal–insulator transition [17].

Figure 1(d) shows the Kohler's analysis of the resistivity curves at varies temperatures. According to Kohler's theory [18]:

$$\frac{\Delta\rho(T,B)}{\rho(T,0)} = F\left(\frac{B}{\rho(T,0)}\right), \qquad (1)$$

where $\rho$ is resistivity, $T$ is temperature and $B$ is magnetic field. If the carrier density of the system is robust to temperature variation, the MR measured at different temperatures can be scaled into a single curve. Here, we assume the function $F\left(\frac{B}{\rho(T,0)}\right) = A(T)\left(\frac{B}{\rho(T,0)}\right)^2$ in the perpendicular field with parameter $A(T)$, due to the parabolic dependence of magnetic field at different temperatures when B // c axis. As shown in Fig. 1(d), the scaled MR curves violet the Kohler's rule with $A(T)$ changes with temperature, indicating that the temperature effect needs to be analyzed.

Figure 2 shows normalized MR (ρ(B)/ρ(0)) of WTe$_2$ measured in different field directions at varied temperatures. In the perpendicular field (B // c axis) configuration, as shown in Fig. 2(a), pronounced Shubnikov-de Haas (SdH) oscillations are clearly visible around 5 T at 2 K. The MR reaches as high as 1,132,200% at 14.7 T and 2 K, about three times larger than the values reported before [9, 12]. With further increasing the temperature, MR decreases and the SdH oscillations are suppressed above 15 K. The field dependences of MR in the parallel field (B // b axis) at different temperatures are plotted in Fig. 2(b). The MR at low temperatures shows a classic parabolic dependence of magnetic field with SdH oscillations at high fields. Compared to the MR measured in perpendicular field, the MR effect measured with transverse field (B // b axis) is decreased to $9 \times 10^4$ % at 14.7 T at 2 K. Surprisingly, when the magnetic field is parallel to the applied current (B // a axis), the MR shows a linear behavior rather than the parabolic behavior [Fig. 2(c)]. The large non-saturated linear MR is achieved as high as 1200% at $T$ = 2 K and $B$ = 15 T. Moreover, the LMR property sustains with increasing temperature until 50 K.

In order to analyze the SdH oscillations measured in perpendicular field (B // c axis) and transverse field (B // b axis), we subtracted the polynomial background in order to show the oscillatory component $\Delta\rho = \rho(T,B) - \rho(T,0)$ [Fig. 3]. The frequency of the SdH oscillations was extracted from fast Fourier transform (FFT) analysis. For the perpendicular case, as shown in the inset of Fig. 3(a), there are four major peaks at 89.5 T, 119.4 T, 141.8 T and 156.7 T, which may stand for four Fermi pockets δ, α, β, γ [12]. The higher harmonics of α peak is also observed. With increasing temperature, the δ peak and γ peak as well as the higher harmonics of α peak are gradually diminished while α peak and β peak remain. Such results indicate that there exist two robust Fermi pockets, which is consistent with the previous study [12]. For the parallel field case (B // b axis), shown in Fig. 3(b), only two obvious frequencies λ and η at 29.8 T and 218.8 T can be observed. We did first principle calculation (see Supplemental Material) for $WTe_2$ and found that there are only two comparable Fermi pockets which are identified as a pair of electron and hole pockets along Γ–X direction. These comparable pockets result in similar charge carrier density for the electrons and the holes, which is robust upon using different exchange-correlation functions within the calculation of density-functional theory, consistent with the earlier theoretical studies [11]. The association of the other two frequencies to the pockets not appearing in the band diagram might be attributed to higher order correlation interactions, which could be an interesting issue for future studies.

To further investigate the anisotropic Fermi surface in $WTe_2$, it is necessary to measure the angular-dependent MR behavior and therefore analyze the evolution of Fermi surface. Figure 4(a) shows MR behavior in the longitudinal rotation by varying magnetic field directions from *c* axis to *b* axis. The MR effect as well as the SdH oscillation is gradually weakened by rotating the magnetic field away from *c* axis. Figure 4(b) demonstrates how the MR behavior changes in the transverse rotation when the magnetic field is tilted from transverse (*b* axis, $B \perp I$) to longitudinal (*a* axis, $B \parallel I$) case. While MR is mostly quadratic changing with fields when B parallel to *b* axis (0 deg), it gets smaller as B rotated away from *b* axis. When the magnetic field is parallel to the applied current direction (B // *a* axis), the MR exhibits a non-saturating but linear behavior, showing an extremely anisotropic characteristic of $WTe_2$.

To better understand the angular-dependent MR in $WTe_2$, we performed transport measurements to investigate the tilting-angle dependence of MR at 2 K at different magnetic fields [Fig. 4(c) and 4(d)]. It is well-known that layered or two dimensional materials are only sensitive to the perpendicular component of the magnetic field $Bcos\theta$, where $\theta$ is the angle between magnetic field and *c* axis of the material. As shown in Fig. 4(c), the observed angular dependence of MR in $WTe_2$ can also be fitted by $|cos\theta|$ function, indicating a quasi-two dimensional nature of the semimetal $WTe_2$. Furthermore, the angular dependent MR oscillations are observed at high magnetic fields in the longitudinal rotation situation (rotation axis *a*) (Fig. 4(c)), while absent in the transverse rotation case (rotation axis *c*) (Fig. 4(d)). When the magnetic field is lower than 6 T, the oscillations disappear, consistent with the observed SdH oscillations in $WTe_2$.

In short, the MR behavior of $WTe_2$ system shows three distinct features: (i) the transverse MR

under large magnetic field show quadratic dependence without saturation [9]; (ii) the violation of Kohler's law for transverse MR; (iii) an exotic longitudinal LMR. The first feature origins from the absence of Hall effect in systems with equal numbers of electron and hole carriers, the transverse MR quadratically increase with the magnetic field without saturation [9, 18]. As to the second one, in the large field limit with $\mu(T)\text{B} \gg 1$, the transverse MR can be written in a simple form [18]: $\frac{\Delta\rho(B,T)}{\rho_0(T)} = [\mu(T)\text{B}]^2 = \left[\frac{B}{n(T)e\rho_0(T)}\right]^2$, here $\mu(T)$ and $n(T)$ denotes the mobility and total carrier density of the system at temperature $T$. Normally, the carrier density $n(T)$ is not sensitively dependent on the temperature, which results in a scaling plot of transverse MR—the Kohler rule [18]. However, in WTe$_2$ single crystal with both electron and hole carriers, the formation of excitons can change the carrier density, thus the Kohler rule can be violated. As shown in Fig. 1(d), the slope of scaling plot of MR firstly increases and then decreases around 20 K when raising the temperature. This behavior indicates that the carrier density is also non-monotonically dependent on the temperature—firstly decreasing then increasing. We discuss the possible mechanism of carrier density change in the context of exciton formation. In WTe$_2$ single crystal, the electron and hole pockets separately lie along the $\Gamma - X$ line. The formation of excitons in three dimensional systems needs the prefect nesting of electron and hole pockets. This severe condition can hardly be satisfied due to different shapes of electron and hole Fermi surfaces. With increasing the temperature, a typical phonon with wave vector $\vec{k}_0$ and phonon frequency $\omega(\vec{k}_0)$ appears, which connects the electron and hole pockets. This typical phonon mode connects the electron and hole pockets, and promotes the formation of excitons. The exciton density is proportional to the phonon number $N[\omega(\vec{k}_0)]$, and the residual free charge carrier density decreases with increasing temperature. This formation of excitons is similar to the liquid-gas transition in previous studies [19, 20]. From the phonon spectrum, we find the phonon energy for $\omega(\vec{k}_0)$ lies in the range of 10 K~ 20 K (See Supplemental Material). With increasing the temperature, thermal activation can destroy the exctions. The free charge carrier density reads $n(T) = \frac{n_0}{\exp(\Delta/k_B T)+1}$ for $k_B T \geq \hbar\omega(\vec{k}_0)$ with $\Delta$ denoting the excitonic gap. Thus, the charge carrier density turns to increase with increasing the temperature. Overall, the phonons assist the exciton formation, and the thermal activation destructs the excitons. These double effects give rise to the non-monotonic carrier density dependence on temperature and result in a minimum charge carrier density around 20 K in Fig. 1(d). Thirdly, we consider the origin of longitudinal LMR at low temperature. Classically, the magnetic field only influence the carrier motion perpendicular to the field, and the longitudinal resistance has no field dependence. However, the quantum effect gives rise to dispersive quasi-one dimensional (quasi-1D) states along the magnetic field with Landau levels in the perpendicular plane. In the quantum limit, the large degeneracy of Landau levels leads to correlated effects. A well-known example is the CDW instability in Bismuth in the quantum limit [21, 22]. However, different from Bismuth, which mainly possesses hole carriers, $WTe_2$ has equal number of electron and hole carriers. The magnetic field B // $a$ axis leads to quasi-1D electron and hole states along $\Gamma - X$ line, and the density for each quasi-1D states satisfies $dN(n,k_x)/dk_x = eB/2\pi\hbar c$. From the Fermi golden rule, the scattering rate between one electron and one hole Landau Level is proportional to the magnetic field. This mechanism may contribute to the observed longitudinal LMR. Moreover, the nesting condition for quasi-1D electron and hole states are much easier to be satisfied than the 3D states. Further microscopic studies are expected to uncover the scattering and the exciton formation in this double carrier correlated system.

In summary, intriguing quantum transport properties in $WTe_2$ are revealed by systematic angle-dependent MR measurements at different temperatures. A large RRR in zero field and metal–insulator transition in perpendicular field are observed. Besides, it is found that the MR shows extremely anisotropic properties. The parabolic MR reaches as high as $1.1322\times10^6$ % at 14.7 T at 2 K when B // $c$ axis but is decreased to $9\times10^4$ % when B // $c$ axis. Surprisingly, an exotic large LMR as high as 1200% at 15 T is observed when B // $c$ axis. An empirical theoretical model of double carrier correlated system is developed based on the balanced hole-electron condition in $WTe_2$ to explain the observations. We expect that our research will encourage further theoretical and experimental studies on this exciting double carrier correlated layered material and corresponding new quantum properties.

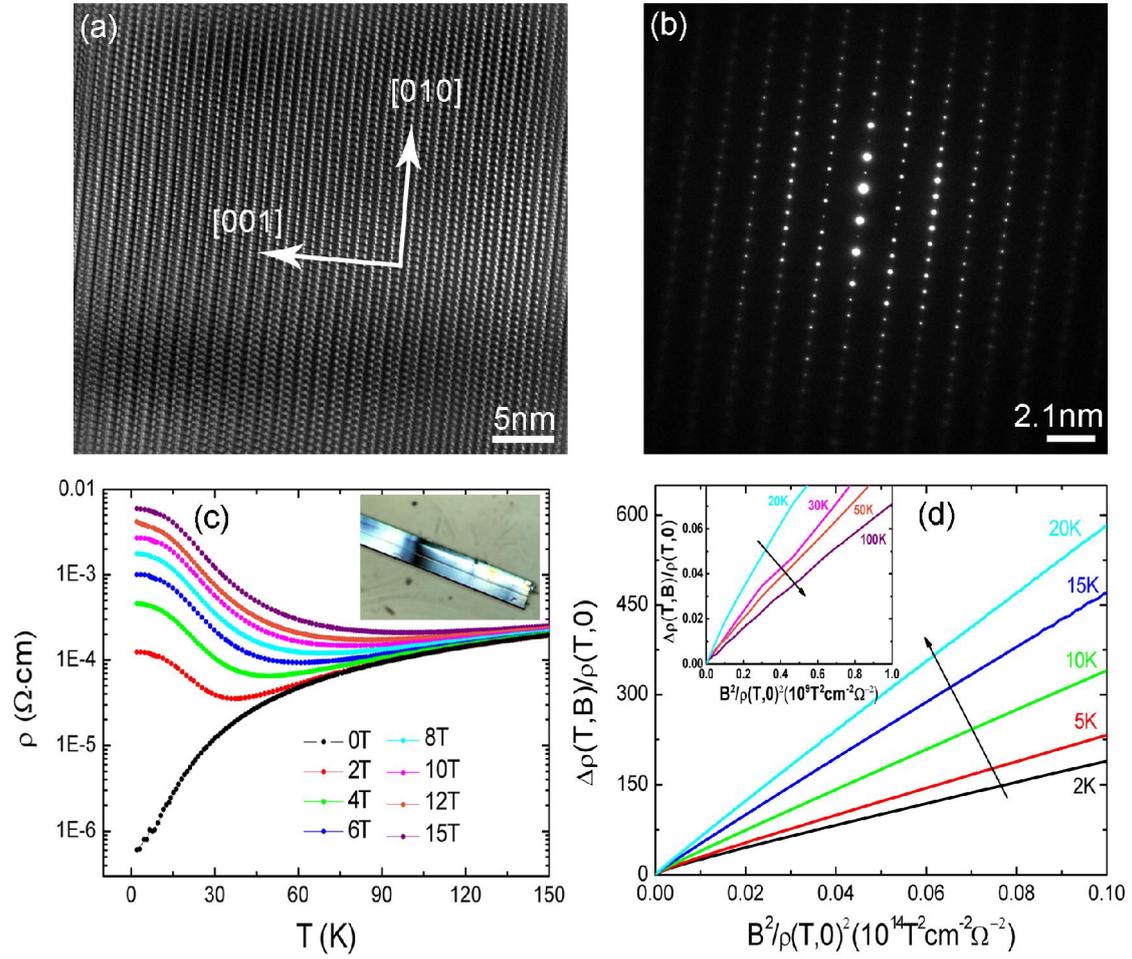

FIG 1. (color online). (a) High-resolution transmission electron microscopy image of WTe$_2$ single crystal. (b) Electron diffraction image looking down the [100] zone axis showing the reciprocal lattice of WTe$_2$. (c) Temperature dependence of longitudinal ρ of WTe$_2$ at different magnetic fields. Inset: the optical image of WTe$_2$ single crystal. (d) Kohler's rule by plotting the ρ($T$,B)/ρ($T$,0) vs. B$^2$/ρ($T$,0)$^2$ from 2 K to 20 K. Inset show the high temperature regime from 20 K to 100 K.

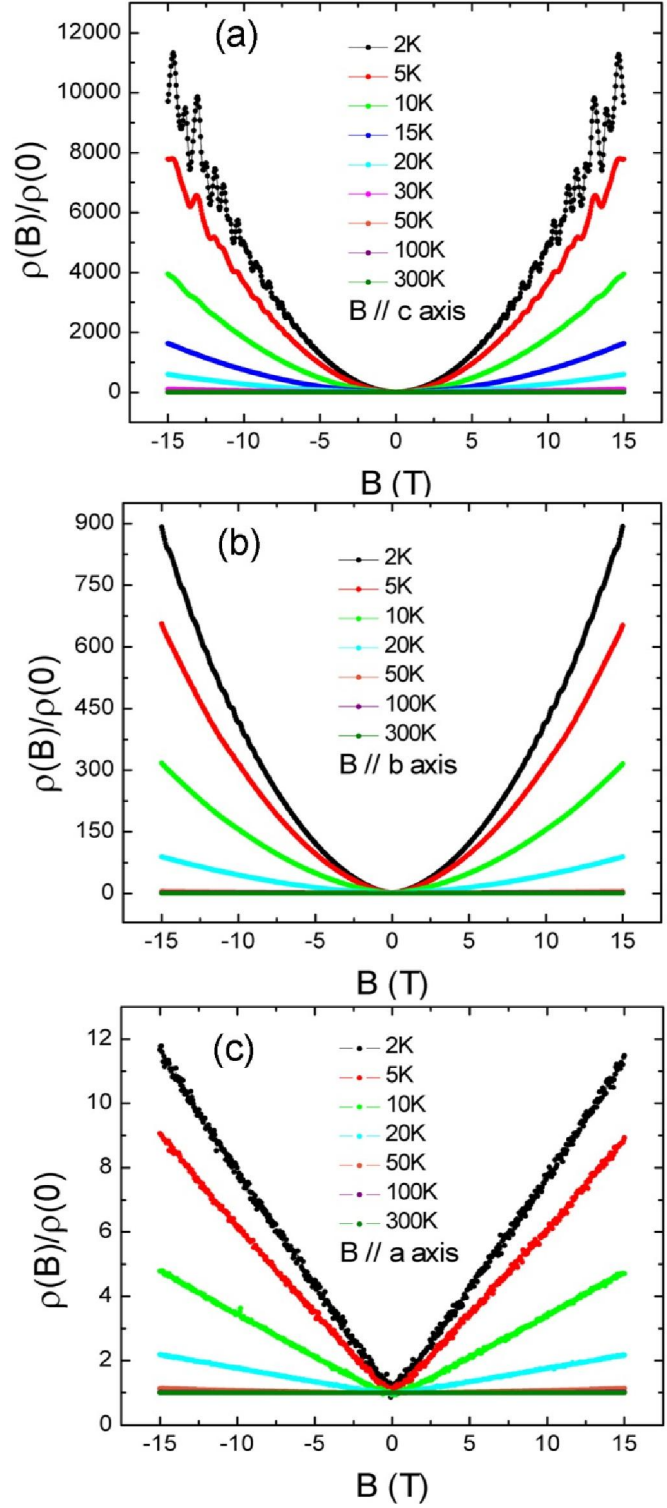

FIG 2. (color online). Normalized magnetoresistivity ρ(B)/ ρ(0) measured in (a) the perpendicular field (B // *c* axis), (b) the parallel field with B // *b* axis and (c) B // *a* axis at different temperatures, respectively.

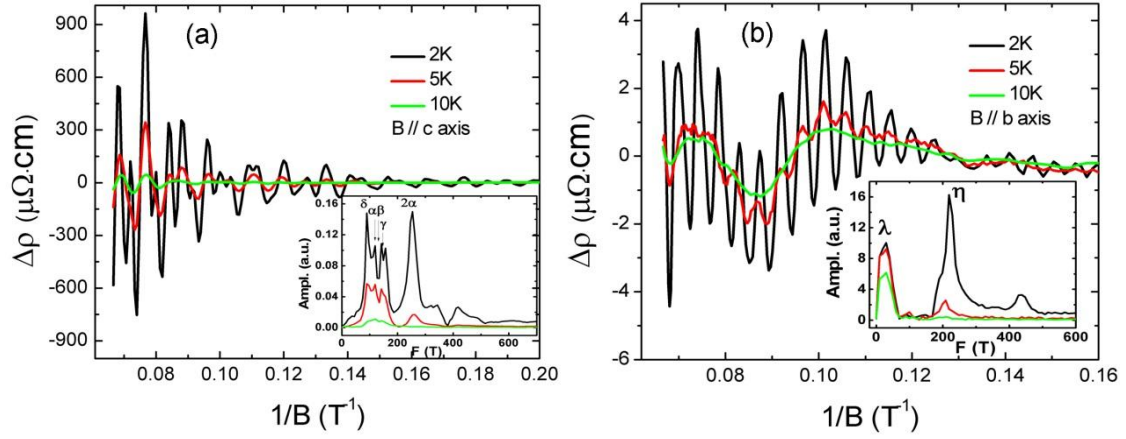

FIG 3. (color online). (a) After subtracting the polynomial background, the SdH oscillatory component Δρ in the perpendicular field at varies temperatures. (b) The oscillatory parts of Δρ in the parallel field with B // b axis. Insets are the corresponding FFT analysis.

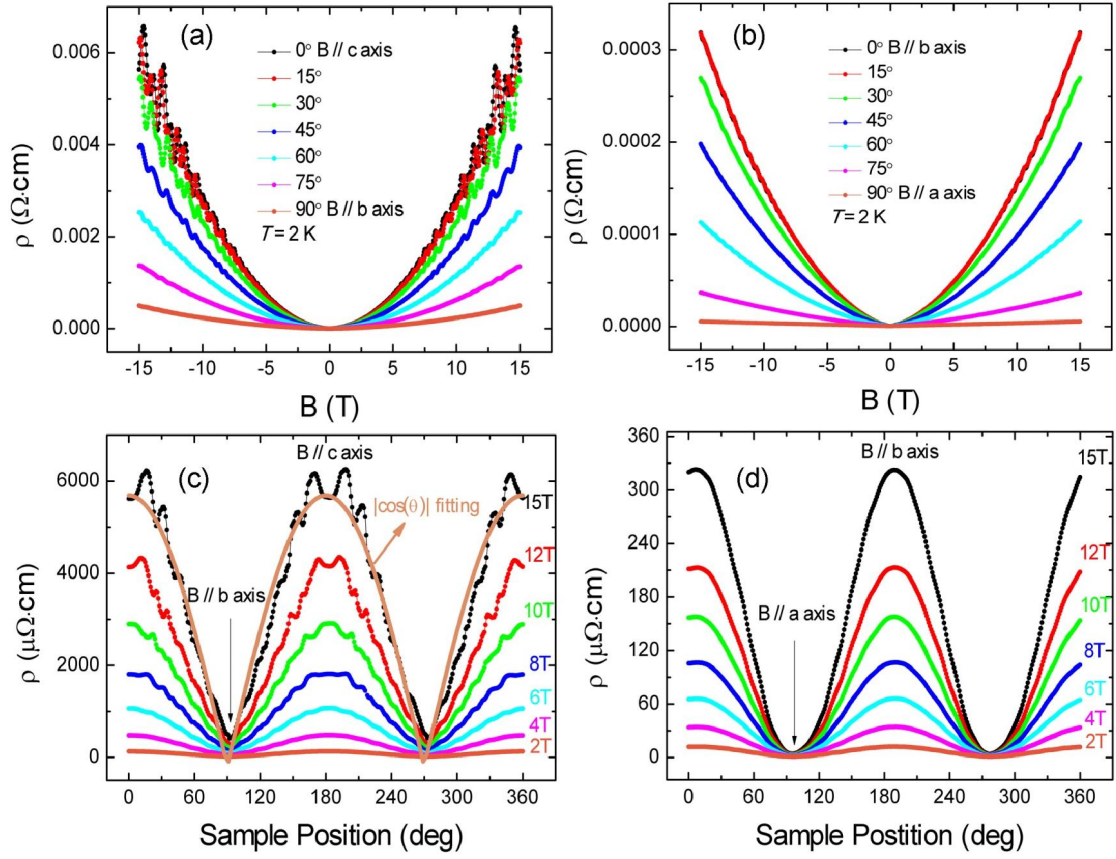

FIG 4. (color online). (a) Field dependent magnetoresistivity at $T$ = 2 K with the magnetic field rotating from $c$ axis (0 deg) to $b$ axis (90 deg) (b) Magnetoresistivity measured at different angles in $ab$ plane at $T$ = 2 K (c) and (d) Magnetoresistivity as a function of tilt angle in different magnetic fields at $T$ = 2 K for longitudinal rotation (rotation axis $a$) and transverse rotation (rotation axis $c$), respectively. The function |cosθ| fitting is shown as solid dark orange curve in Fig. 4(c).

Acknowledgements

We acknowledge Qian Niu and Hua Chen for helpful discussions. This work was financially supported by National Basic Research Program of China (Grant Nos. 2013CB934600 and 2012CB921300), the National Natural Science Foundation of China (Nos. 11222434, 11174007), and the Research Fund for the Doctoral Program of Higher Education (RFDP) of China. DGM acknowledges support from the Gordon and Betty Moore Foundation's EPiQS Initiative through Grant GBMF4416. JQY acknowledges support from the US Department of Energy, Office of Science, Basic Energy Sciences, Materials Sciences and Engineering Division.

# Supplementary Material

## Anisotropic Magnetotransport and Exotic Longitudinal Linear Magnetoresistance in WTe$_2$ Crystals

Yanfei Zhao[1,2], Haiwen Liu[1,2], Jiaqiang Yan[3,4], Wei An[2,5], Jun Liu[6], Xi Zhang[1,2], Hua Jiang[7], Qing Li[8], Yong Wang[6], Xin-Zheng Li[2,5], David Mandrus[3,4], X. C. Xie[1,2], Minghu Pan[9],* and Jian Wang[1,2]*

[1] International Center for Quantum Materials, School of Physics, Peking University, Beijing 100871, China
[2] Collaborative Innovation Center of Quantum Matter, Beijing 100871, China
[3] Department of Materials Science and Engineering, University of Tennessee, Knoxville, Tennessee 37996, USA.
[4] Materials Science and Technology Division, Oak Ridge National Laboratory, Oak Ridge, Tennessee 37831, USA.
[5] School of Physics, Peking University, Beijing 100871, China
[6] Center of Electron Microscopy, State Key Laboratory of Silicon Materials, Department of Materials Science and Engineering, Zhejiang University, Hangzhou, 310027, China
[7] College of Physics, Optoelectronics and Energy, Soochow University, Suzhou 215006, China
[8] Institute of Functional Nano and Soft Materials (FUNSOM) and Collaborative Innovation Center of Suzhou Science and Technology, Soochow University, Jiangsu 215123, China.
[9] School of Physics, Huazhong University of Science and Technology, Wuhan 430074, China.

*Correspondence and requests for materials should be addressed to M. P. (email: mhupan @gmail.com ) or to J. W. (jianwangphysics@pku.edu.cn)

**Contents:**

I.  First-principles calculation on the band structure

II.  Fermi surface demonstration

III.  Phonon dispersion

**I. First-principles calculation on the band structure:**

The peculiar property of the WTe2 is mainly due to its balanced electron and hole pockets in the Brillouin zone (BZ). The Shubnikov-de Haas (SdH) oscillation and angle resolved photoemission spectroscopy (ARPES) experiments have claimed that there are four pockets and two of them are robust under the variation of temperature and pressure. However, previous calculation by Xu *et al.*[1] shows only two pockets with another two potential pockets.

We, therefore, also perform the same calculation to check it within the framework of density-functional theory (DFT), adopting the full-potential linearized augmented plane wave (FP-LAPW) WIEN2k package with the PBE exchange-correlation functional.[S1] The self-consistent calculation of the $WTe_2$ bulk was performed based on the experiment structure[S2] with the $R_{MT}K_{max}$ equals 8 and the 10000 k points sampling in BZ. The spin-orbital coupling (SOC) was considered using a second-variation method.

The band dispersion is then calculated along the high symmetry k point path which is kept the same as in the work of Xu *et al.*.[1] As shown in the Fig. S1 (a) and (c), the band structure calculated using PBE functional reproduces the results in Ref. 1, *i.e.* there are only two pockets close to the Fermi surface. And by including the SOC, there is in general very little difference between (a) and (c).

We then tried a newly developed functional namely modified Becke-Johnson (mBJ) potential, to redo the band structure calculation.[S3] The mBJ functional can serve as a good substitution for the more accurate *GW* method in which higher order correlation effects are considered in many cases. We use the one-shot mBJ approach, the detail of which can be found in the Ref. S4. Fig. S1 (b) and (d) show the band structure using mBJ without and with SOC, respectively. Without SOC, the band dispersion of electron and hole pockets near the Fermi surface becomes entangled which is inconsistent with the experiment. By including the SOC, this is rectified. Compared to the PBE+SOC results, the band dispersion obtained with mBJ+SOC shows two little differences: a) away from the Fermi surface, the gap between the conduct and valence band enlarged. b) near the Fermi surface, the semi-metal property, in other word, the area of the two pocket is enhanced. Therefore, mBJ calculation cannot give four pockets either and the two original potential U-Z pocket become less obvious. For insulator, the mBJ can be a good approach to evaluate band gap and often an enlarged gap can be obtained compared with PBE. However, the semi-metal behavior may lead the mBJ calculation to wrong direction. In short, we found that the two pockets in the Γ-X direction are quite robust under different functionals. The computational evidence of the other two pockets observed from experiment may have to resort to more delicate method, such as GW or Quantum Monte Carlo.

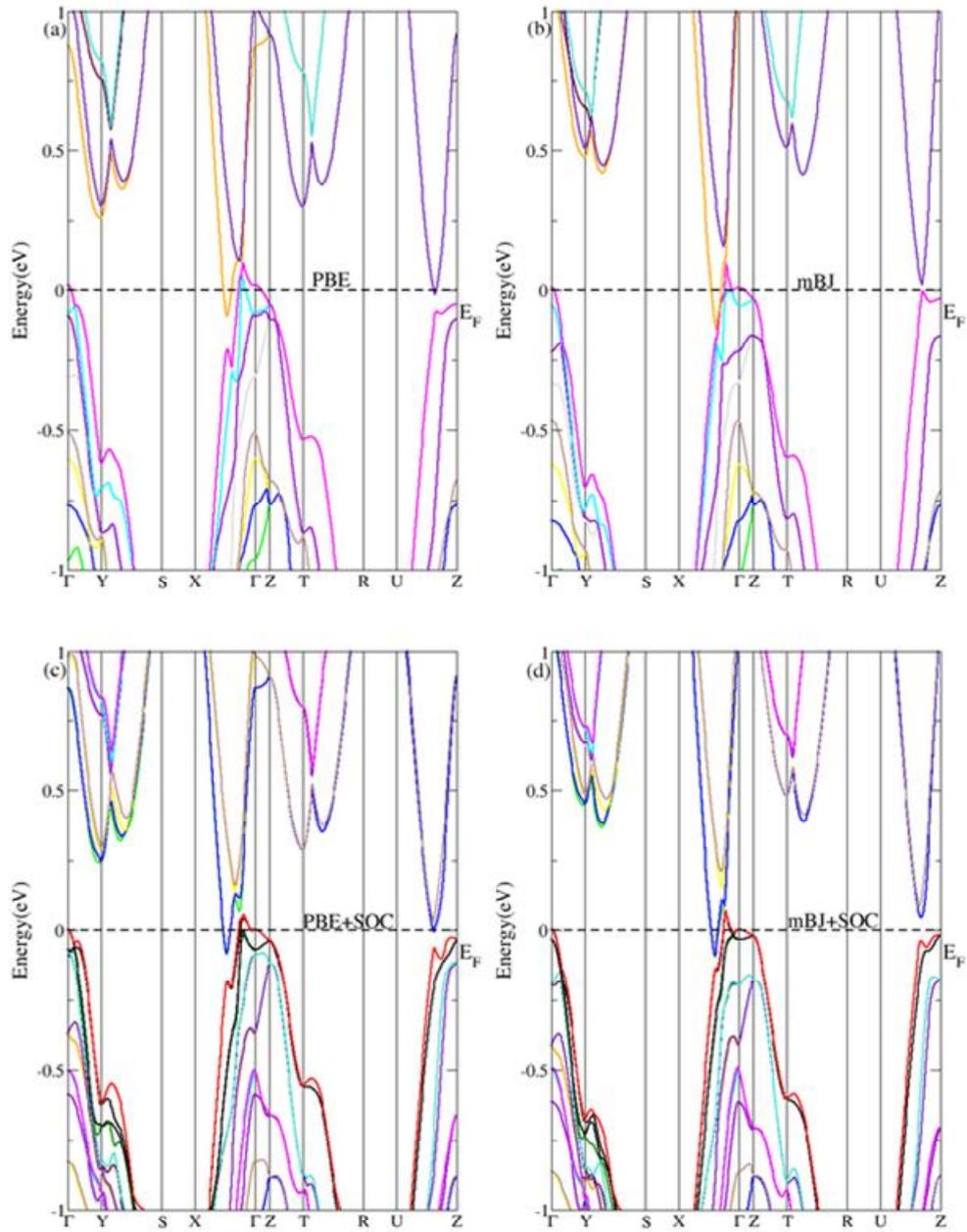

**Figure S1 | Calculated Band structure for the WTe$_2$ bulk with different functional and with/without SOC effect.** (a) PBE without SOC, (b) mBJ without SOC, (c) PBE with SOC and (d) mBJ with SOC.

## II. Fermi surface demonstration:

The extremely anisotropic MR revealed by the transport experiment has been attributed to the anisotropic Fermi surface. To better understand the relationship between the MR property and the Fermi surface, we reproduce the 3D Fermi surface with SOC effect taking PBE as the functional and visualized using the Xcrysden program. As shown in Fig. S2, the overview of the 3D Fermi surface and also different projection plane is demonstrated. The yellow pocket stands for the electron, the grey one for the hole. It shows that there are two pockets lying in the Γ-X direction. Based on the three projection plane, the cross-section area of the Fermi surface from different perspective varies a lot. This strong anisotropy shown by the Fermi surface leads to the peculiar MR property.

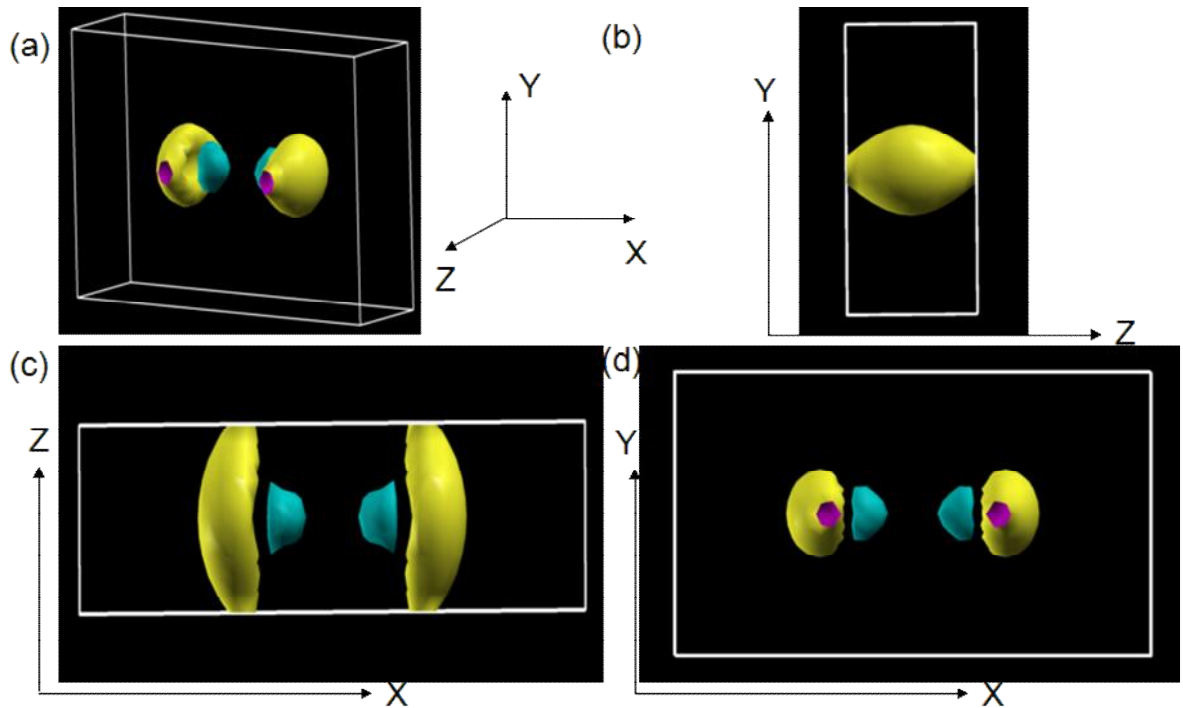

**Figure S2 | Calcutated 3D Fermi surface for WTe$_2$ bulk showing electron (yellow) and hole (turquoise) pockets. (a) an overview of the 3D Fermi surface. (b),(c),(d) the projection of the Fermi surface at the Y-Z, Z-X and Y-X plane.**

### III. Phonon dispersion:

As mentioned in the main article, one of the possible reasons for the violation of Kohler's law for transverse MR could be the formation of the excitons. Similar to the case in the liquid-gas transition, the phonons in the range of 10~20K would mainly contribute to this mechanism.

To verify our conjecture, we performed a phonon calculation within the first-principles DFT framework using the CASTEP package.[S5] The structure of $WTe_2$ bulk was first geometrically optimized based on the experiment structure using the PBE functional at the parameter set which ensures the convergence of the calculation: Energy cutoff is 500 eV, the k-mesh is 4x2x1 and the relaxation threshold is 0.01eV/Å. The optimized lattice parameter is 3.47, 6.26 and 16.54 Å for a, b and c, respectively, which is comparable to the 3.48 6.25 and 14 Å for the experiment.

Based on the optimized structure, the phonon dispersion calculation was performed using the linear response method, the result is shown in the Fig. S3, which is in general consistent with the previous calculation.[S6] In order to compare between the band structure and the phonon dispersion, the k-path was set to be the same as in the band structure. We can see clearly that there are acoustic modes at the range of 10~20K, residing in the Γ-X direction, which may assist the formation of excitons.

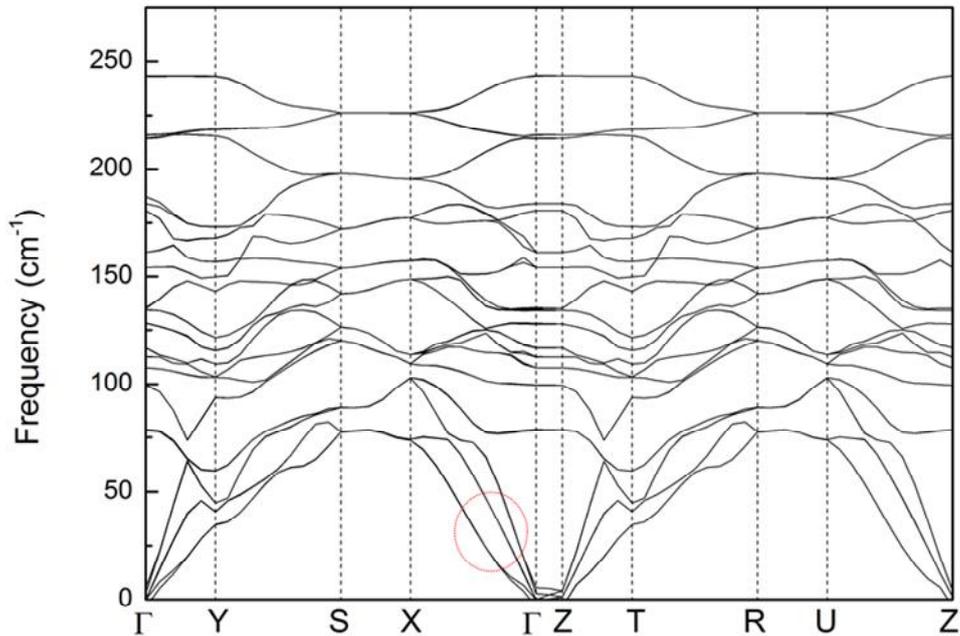

**Figure S3 | Phonon dispersion curve obtained from the first-principles calculation along the high symmetry points of the orthorhombic Brillouin zone for $WTe_2$ bulk, the dashed red circle is the region where the phonon which may contribute to the**

**exciton formation.**